\documentclass[fleqn,twoside]{article}
\usepackage{espcrc2}

\usepackage{graphicx}
\usepackage{epsfig}
\usepackage[figuresright]{rotating}

\newcommand{\AmS}{{\protect\the\textfont2
  A\kern-.1667em\lower.5ex\hbox{M}\kern-.125emS}}

\newcommand{\mincir}{\raise -2.truept\hbox{\rlap{\hbox{$\sim$}}\raise5.truept
\hbox{$<$}\ }}
\newcommand{\magcir}{\raise -2.truept\hbox{\rlap{\hbox{$\sim$}}\raise5.truept
\hbox{$>$}\ }}
\def\hmpc{\ifmmode\,h^{-1}\,{\rm Mpc}\;\else$h^{-1}\, {\rm Mpc}\;$\fi}
\def\kmpc{\ifmmode\,h\,{\rm Mpc^{-1}}\;\else$h\,$Mpc^{-1}\;\fi}

\def\kmpc{\rm \,h\,Mpc^{-1}}

\def\xip{$\xi(r_p,\pi)$\ }
\def\xip{$\xi(r_p,\pi)$\ }

\def\pk{{$P(k)$}\ } 

\def\kms{\,{\rm km\,s^{-1}}}

\def\n_med{{\left<n\right>}}

\title{Recent Advances in Large-Scale Structure and Galaxy Formation
Studies
} 

\author{L. Guzzo\address[MCSD]{INAF - Osservatorio Astronomico di Brera\\
	Via Bianchi 46, I-23807 Merate (LC), Italy}%
        \thanks{e-mail: guzzo@merate.mi.astro.it}
}
       
\begin{document}

\begin{abstract}
I review recent progress in the study of the large-scale structure
of the Universe, covering the following areas: (1) Results from
recently completed or ongoing redshift surveys of galaxies and
X-ray clusters; (2) Measurements of the power spectrum of fluctuations
approaching Gpc scales; (3) Redshift-space distortions and their
cosmological use; (4) Structure at high redshifts and its connection
to galaxy formation.

\vspace{1pc}
\end{abstract}

\maketitle

\section{INTRODUCTION}

It is a particularly fortunate moment to review the field of
large-scale structure\footnote{Review to appear in {\it Topics
in Astroparticle and Underground Physics - TAUP2001}, (LNGS, September
2001), Nucl.Phys. B, A. Bettini et al. eds., Elsevier}, in the
light of the impressive series of results that appeared during the
last year or so, as a consequence of the (entire or partial)
completion of large surveys of galaxies and clusters of galaxies. The
enthusiasm for the new large-scale structure measurements has been
further animated by the immediate possibility to compare them to the
recent fundamental results obtained by microwave background
experiments, which measured with unprecedented precision the power
spectrum of anisotropies over scales finally overlapping those probed
by galaxy surveys (see De Bernardis, this volume).  In this brief
non-specialist review I have tried and assess this enthusiasm, in the
spirit of providing a general, though clearly incomplete, guide to
what seem to me the most promising recent results on large-scale
structure in the local Universe.  Some emphasis is placed on the
importance of understanding the ``bias'', i.e. the relation between
the distribution of the objects under study and the matter whose
gravity governs the overall evolution of clustering.  Finally,
prospects for studying such evolution through deep surveys of galaxies
are quickly touched in the last section.

The recent flow of results makes it even more
difficult for such a brief review to be complete, or at least
balanced. I therefore apologize to those colleagues whose work was
inadvertently overlooked.  The time scale is tuned to September 2001,
but I have added a few references until January 2002 which seemed
helpful for a clearer picture. 

\section{COSMOLOGICAL FRAMEWORK}

Let us set the basic scene for interpreting the observations we shall
discuss here.  It will be interesting, at the end, to verify whether
the assumed framework is corroborated by the latest results.

The current ``standard'' model for the origin and evolution of
structure in the expanding Universe is the Cold Dark Matter (CDM) model
\cite{CDM}, whose global features provide a framework which is
remarkably consistent with a large number of observations.  The ``Cosmology
2000'' version of the model, which takes into account the independent
evidences for a flat geometry (from the angular power spectrum of
anisotropies in the Cosmic Microwave Background \cite{CMB}) and an
accelerated expansion (from the luminosity-distance relation of
distant supernovae, used as ``standard candles'' \cite{SNIa}) is one
where CDM, in the form of some kind of weakly-interacting
non-relativistic particles (see pertinent articles in this volume),  
contributes about 30\% of the total density, with the remaining 70\%
provided by a ``dark energy'' associated to a {\it
Cosmological Constant}.  I will comment at the end of this review on
how comfortable we should feel in front of the number of ``unseen''
ingredients of this model.  Here we shall use the model as it is, in
fact ``just a model'', i.e. a physically motivated machinery which
works remarkably well when confronted with a variety of observations.

\begin{figure}
\centering
\epsfxsize=6cm 
\epsfbox{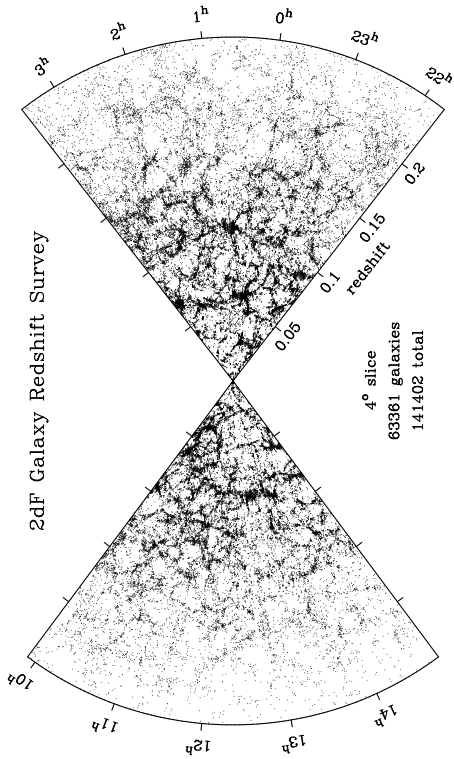} 
\caption{The distribution of over 63,000 galaxies in two 4-degree thick
slices extracted from the total of more than 210,000 galaxies that
currently make up the 2dF Galaxy Redshift Survey (2dFGRS, figure from
\cite{2dF_xipz}).
}
\label{2dF_cones}
\end{figure}
Choosing CDM (or any other model) means specifying a {\it Transfer
Function} $T(k)$.  This can be thought of as describing a linear
amplifier\footnote{$k$ is the Fourier 
wavenumber, i.e. the inverse of a 3D spatial scale $\lambda=2\pi/k$,
measured in $\hmpc$, with $h$ being the Hubble constant in units of
$100 \hmpc$.  Most recent determinations indicate $h\simeq 0.7$ with
about 10\% error, see W. Freedman contribution to this volume.} which
filters the primordial spectrum of fluctuations (possibly of the
scale-invariant form $P_o(k)\propto k$ generally predicted by
inflation) to produce the shape of power spectrum we can still observe
today on large [$k\mincir 2\pi/(10 \hmpc) $] scales, $P(k) = |T(k)|^2
P_o(k)$ \cite{JAP,Paddy}.  One of the nice features of the CDM spectral
shape in any of 
its variants is to naturally lead to a {\sl hierarchical} growth of
structures, where larger entities are continuously formed from the
assembly of smaller ones \cite{White_Rees}.  Within the {\sl gravitational instability}
picture, the formation of galaxies and larger structures is completely
driven by the gravitational field of the dark matter, with our
familiar {\sl baryonic} matter representing only a tiny bit of the
mass ($\sim 2\%$ of the total energy density).  The lighting-up of
galaxies and other luminous objects depends then on how the baryons
cool within the dark matter haloes and form stars, ending up as 
the only directly visible peaks of a much larger, invisible structure.

This increasing complexity in the physics involved is reflected by the
limits in the predicting power of current detailed models of galaxy formation.
Predictions from purely gravitational n-body experiments concerning
the overall clustering of the dark mass can be regarded as fairly robust
\cite{HubbleVolume}.  More complex semi-analytical calculations
addressing the history of galaxy formation have seen exciting progress
during the last few years \cite{Somerv,Kauf,Durh,GRASIL}, but they clearly
still depend on a large number of physically reasonable but
``tunable'' parameters.

Direct measurements of large-scale structure are a classical
test-bench for CDM models and they have, for example, been the reason
for rejecting the original Einstein-DeSitter ($\Omega_{Matter}=1$)
version of the
model, whose transfer function is inconsistent with the observed
balance of large- to small-scale power \cite{APM90}.  The main
problem in this game is that most observations of large-scale
structure necessarily need to
use radiating objects as tracers of the mass distribution, and thus
need to go through the uncertainties mentioned above to allow
meaningful comparison to model predictions \cite{BG2001}.

\section{PROGRESS IN LARGE-SCALE STRUCTURE STUDIES}

\subsection{Galaxy Redshift Surveys}

Since the 1970's, redshift surveys of galaxies are the primary way to
reconstruct the 3D topology of the Universe \cite{rood}. Last year
has seen the completion and public release \cite{2dF_release}
of the first 
100,000 galaxy redshift measurements by the Anglo-Australian 2dF
Galaxy Redshift Survey\footnote{http://www.mso.anu.edu.au/2dFGRS}, the
largest complete sample of galaxies with measured distances to date
\cite{2dF_release}.  This survey includes all galaxies with blue
magnitude $b_J$ brighter than $\sim 19.5$, mainly over two areas covering
$\sim 2000$ square degrees in total, to an effective depth of about
$600 \hmpc$ ($z\sim 0.2$).  The immediate precursors of this survey
\cite{LCRS,ESP} 
reached a similar depth, but over much smaller areas: for comparison,
the Las Campanas Redshift Survey (LCRS \cite{LCRS}), completed in 1996,
measured a total of 
16,000 redshifts, compared to the 250,000 that will eventually form
the full 2dF survey.  A 
plot of the galaxy distribution within the two main sky regions of
this survey is shown in Fig.\ref{2dF_cones}.  Here one can appreciate 
in detail the wealth of structures typical of the distribution of
galaxies: clusters, superclusters (filamentary or perhaps sheet-like)
and regions of very low density, the {\sl voids}
\cite{rood}.  A number of important results have been published in
2001 and are continuing to appear from the first completed part of the
survey. I will discuss some of them in more detail in the following
sections (see also \cite{JAP_texas} for an overview).

In a parallel effort, the Sloan Digital Sky Survey
(SDSS)\footnote{http://www.sdss.org/} is on its way covering a
large fraction of the Northern sky with a CCD survey in five
photometric bands ($u^\prime$, $g^\prime$, $r^\prime$, $i^\prime$,
$z^\prime$), while in parallel measuring redshifts for one million
galaxies over the same area \cite{SDSS_overview}.  The SDSS photometric
survey reaches a typical red magnitude $r^\prime \sim 23$, with the
galaxy redshift survey being limited to $r^\prime = 17.7$.  The SDSS
has also released an early fraction of its data
(http://archive.stsci.edu/sdss/), including photometry and
spectroscopy over 462 square degrees.  There is no doubt that
this represents the largest and most comprehensive galaxy survey work
ever conceived.  The multi-band photometric survey is going to be of immense
value for a number of studies, as estimating {\sl photometric
redshifts} \cite{AFSoto,Yee} to much larger depth than the direct 
spectroscopic survey, or select sub-samples of objects with
well-defined colour/morphology properties.  One relevant example of
such colour selections has been the discovery of several high-redshift
($z>5$) quasars, including the $z=6.28$ case for which the first
possible detection of the long-sought Gunn-Peterson effect,
essentially the fingerprint of the ``dark-ages'', has been recently
reported \cite{Gunn-Peterson}.  Another important application will be
the selection of about $10^5$ ``red luminous'' galaxies with
$r^\prime<19.5$, that will be observed spectroscopically providing a
nearly volume-limited homogeneous sample of ``old'' galaxies out to a redshift
$z\simeq 0.5$, by which to study the clustering power spectrum on
extremely large scales and its evolution
\cite{Eisenstein_LRG}.  A detailed progress report on the SDSS has
been presented at the meeting by Josh Freeman and the reader is
addressed to the corresponding contribution for references to the
first general analyses from the survey.

Both the 2dF and SDSS redshift surveys rely upon the large multiplexing
performances of fiber-fed spectrographs, that allow the light from
several hundred galaxies over a field of view of 1-2 degrees to be
conveyed into the same slit on the spectrograph.  This specific
technology, in various forms, has been the key to the explosion of the
redshift survey industry in the 1990's, bringing the efficiency from the $10$
redshifts/night for galaxies brighter than blue magnitude $b \sim 14$ of
the 1970's, to the current $2500$ redshifts/night 
to $b \sim 19.5$ (see e.g. \cite{Texas99} for a more
accurate account).

\subsection{Surveys of X-ray Clusters of Galaxies}

Clusters of galaxies have a honourable history as a
complementary tracer of large-scale structure (see
e.g. \cite{Bob_Sesto}).  Especially before the current era, when 
$N>100,000$ galaxy redshifts are becoming available over
comparable volumes, groups and clusters have represented the most
efficient alternative to map very large volumes of the Universe,
exploring in this way the gross structure and its statistical
properties in the weak clustering regime.   
\begin{figure}
\centering
\epsfysize=7.3cm 
\epsfbox{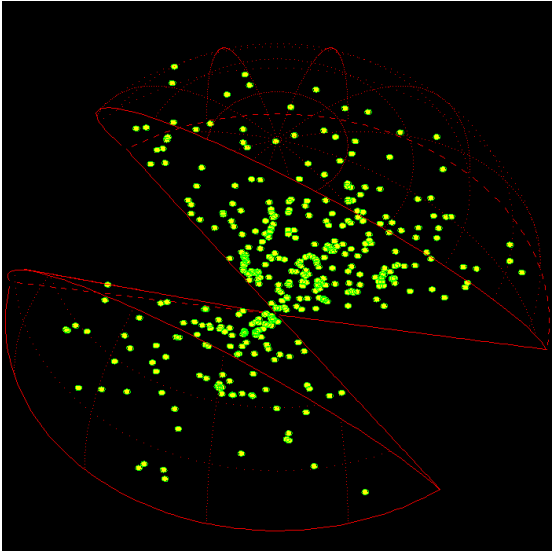} 
\vspace{-1cm}
\caption{The spatial distribution of X-ray clusters in the REFLEX
survey, out to $600 \hmpc$ (from \cite{BG2001}).   Note that here
each point corresponds to a cluster, containing hundreds or thousands
of galaxies.  Structure is here mapped in a coarse way, yet sufficient
to evidence very large structures as the ``chains'' of clusters
visible in this picture. } 
\label{fig:reflex_cone}
\end{figure}

X-ray selection represents currently the most physical way by which to
identify and homogeneously select large numbers of clusters of
galaxies\footnote{A notable  
powerful alternative, so far limited by technical development, is
represented by radio surveys using the 
Sunyaev-Zel'dovic effect.  In this case one measures, in the radio
domain, the CMB spectral distortions produced in the direction of a
cluster by the Inverse Compton scattering of the CMB photons over the
energetic electrons of the intracluster plasma (see
e.g. \cite{Bartlett_Sesto} for a review). } (see also discussion in
\cite{Ellis_Sesto}).  
Clusters shine in
the X-ray sky thanks to the {\it bremsstrahlung} emission produced by
a hot plasma ($kT\sim 1-10$ KeV) trapped within their potential
wells.  The bolometric emissivity (i.e., the energy released per
unit time and volume) of this thin gas is proportional to its density
squared and to $T^{1/2}$.
Such dependence on $n^2$ makes clusters stand out more in the
X-rays than in the optical light 
(i.e. galaxy) distribution ($\propto n$).  

Under the assumption of hydrostatic equilibrium, the intracluster gas
temperature, measured through the X-ray spectrum, is a direct probe
of the cluster mass: 
$kT\propto \mu m_p \sigma_v^2 \sim G\mu m_p M_{vir}/(3 r)$ (where
$m_p$ is the proton mass, $\mu\simeq 0.6$ the gas mean molecular
weight, $\sigma_v$ the galaxy 1D velocity dispersion and $M_{vir}$
the cluster virial mass).  
X-ray luminosity, a more directly observable quantity with current
instrumentation, shows a good correlation with temperature,
$L_{X}\propto T^\alpha$ with $\alpha\simeq 3$ and a scatter
$\mincir 30\%$.  
The practical implication,
even only on a phenomenological basis, is that clusters
selected by X-ray luminosity are in practice mass-selected, with an
error $\mincir 35$ \% (see e.g. \cite{BorganiRDCS2001} and references
therein for a more critical discussion).  Last, but not least,
the selection function of an X-ray cluster survey can be determined to
high accuracy, knowing the properties of the X-ray telescope used, in
a similar way to what is usually done with magnitude-limited samples
of galaxies \cite{Rosati_1}.  This is of fundamental importance if one
wants to compute statistical quantities and test
cosmological predictions as, e.g., the mean density or the
clustering of clusters above a given mass threshold \cite{BG2001}.

Fig.~\ref{fig:reflex_cone} plots the large-scale distribution of X-ray
clusters from the REFLEX (ROSAT-ESO Flux Limited X-ray) cluster
survey, the largest redshift survey of X-ray clusters with homogeneous
selection function to date \cite{REFLEX_survey_paper}.  This data set,
completed in 2000 and publicly released at the
beginning of 2002, is based on the X-ray all-sky survey performed
by the ROSAT satellite in the early 1990's (see
e.g. \cite{Henry,Piero_ARAA} for a comprehensive summary).  
REFLEX includes 452 clusters over the southern celestial hemisphere 
and is more than 90\% complete to a flux limit of 
$3 \times 10^{-12}$ erg s$^{-1}$ cm$^{-2}$ (in the ROSAT energy band, 0.1-2.4
keV). 
The volume explored is larger than that of the 2dF survey and
comparable to the volume that will be filled by the SDSS
1-million-galaxy redshift 
survey\footnote{The SDSS will then probe a much larger volume using
a uniform sample of old (``Luminous Red'') galaxies selected through
their colours out 
to $z\sim 0.5$ \cite{Eisenstein_LRG}.}.  In the cluster distribution
the fine structure visible in Fig.~1 is obviously lost; however, a
number of cluster agglomerates and filamentary structures with large
sizes ($\sim 100 \hmpc$) are clearly evident, showing that
clustering is still strong on such very large scales.

\section{STATISTICAL RESULTS ON LARGE-SCALE CLUSTERING}

\subsection{The Power Spectrum of Fluctuations}

We have seen that large-scale structure models as CDM are specified in
terms of a specific shape for the power spectrum of density fluctuations \pk.
Analogously to standard signal theory, the power spectrum describes
the squared modulus of the amplitudes $\delta_k$ (at different spatial 
wavelengths $\lambda=2\pi/k$) of the Fourier components of the
fluctuation field $\delta = \delta\rho/\rho$ \cite{JAP}.  Studying the
power spectrum of the distribution of luminous objects on sufficiently
large scales, where the growth of clustering is still independent of
$k$, we hope to recover a relatively undistorted information to test
the models.  
The uncertainties in relating the observed \pk of, 
e.g., galaxies to that from the theory are due to (a) nonlinear
effects that modify the linear shape below some scale; (b) the
relation between the luminous tracers for which we are measuring \pk
and the mass distribution, that is what the models predict.  While
the first problem can be circumvented by pushing redshift surveys to
larger and larger scales (and/or following nonlinear evolution through
numerical simulations), the second one involves knowing the
so-called {\it bias} factor (or function). This can be defined
as the ratio between the variances in galaxy counts
and in the mass, $\left({\delta n(r) /
\left<n\right>}\right)_{rms}=b\left({\delta\rho(r) /
\left<\rho\right>}\right)_{rms} \label{eq:bias}$.
A significant amount of work has been dedicated to the problem of
galaxy biasing 
during the last fifteen years, both theoretically
(e.g. \cite{enzo_dekel})
and observationally (e.g. establishing that for ``normal'' optically
selected galaxies we have $b\simeq 1-1.5$ over a fair range of scales
\cite{Benoist}).  However, understanding the physical 
origin of the bias involves comprehending the details of how the
(mainly dark matter) mass of a galaxy governs the visible stellar
light we use to select it for our surveys. 
Despite the great advances of the last few 
years in our knowledge of the early phases of galaxy formation
\cite{Ellis2001}, we are still relatively ignorant about the details
of these processes at early epochs. 
\begin{figure}
\centering
\epsfxsize=8cm 
\epsfbox{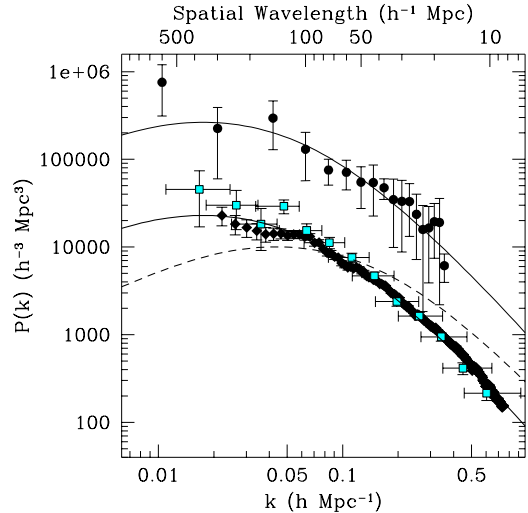} 
  \vspace{-2cm}
\caption{The power spectrum of 2dF galaxies and REFLEX clusters.  {\it
Filled diamonds}: estimate using 147,000 redshifts by the 2dF team
\cite{Percival2001}; {\it open squares}: Tegmark et al. analysis of the
100,000-redshift public release, with accurate treatment of window
aliasing and error covariances \cite{Tegmark_2dF});  {\it filled
circles}:  REFLEX clusters in a $600\hmpc$ box \cite{reflex_pk}.  
{\it Dashed line}: Einstein-De Sitter CDM model; {\it lower solid line}:
Lambda-CDM ``standard'' model (as defined in the first section); both  
are normalized to match the amplitude of CMB fluctuations
\cite{Bunn_White};  {\it upper solid line}:  same Lambda-CDM model, but
renormalized (``biased'') according to the mass selection
function of REFLEX clusters \cite{reflex_pk,BG2001}. 
}
\label{fig:pk}
\end{figure}
As a consequence $b$ remains normally
a free parameter when comparing galaxy clustering to the models.  As I
shall discuss shortly, the situation can be more favourable when
measuring \pk using X-ray selected clusters.

The 2dF and REFLEX surveys have produced the best estimates to date of
the power spectrum of galaxies and X-ray clusters, respectively.
Fig.~\ref{fig:pk} compares these data sets directly.   One can notice
the remarkable similarity of the shape of \pk between galaxies and
clusters.
Such simple proportionality is here seen so clearly thanks to the size
and the quality of these two surveys. This provides a direct confirmation of
the bias scenario, where clusters form at the high, rare peaks of the
mass density distribution \cite{Kaiser84} and for this reason display
a stronger clustering amplitude.  
%
In the same figure I have also plotted the predictions for the power
spectrum of the mass from two variants of the CDM family, computed as
described in \cite{Eisenstein_Hu}.  What we
defined as the ``standard'' model  in the beginning ($\Omega_M\simeq 0.3$,
$\Omega_\Lambda\simeq 0.7$, $h=0.7$)  provides in general an
excellent fit to the 2dF power spectrum, with a bias factor
(i.e. normalization) close to unity\footnote{In fact, once we fix the
primordial spectrum $P_o$, in a pure CDM Universe the observed shape
depends only on $\Omega_M$, not on $\Omega_\Lambda$ (which on the other
hand influences the normalization).  In the literature, this is often
parameterized through a {\sl shape parameter} $\Gamma=\Omega_M\, h\,
f(\Omega_b)$, where $f(\Omega_b)\sim 1$ in case of negligible baryon
fraction.  Our fiducial ``Cosmology 2000'' model, therefore, has
$\Gamma\simeq 0.2$.}.
The upper solid curve, on the
other hand, is the same model re-normalized as
\begin{equation}
P_{clus}(k) = b_{eff}^2 P_{CDM}(k)\,\,\,\,\, ,
\end{equation}
where the effective bias factor $b_{eff}$ has been computed taking into
account the effective mass range of the cluster sample, using a
relatively straightforward theory \cite{MW96,Sheth} (see
\cite{reflex_pk} for more details).  It is for these computations that
a well-understood mass selection function of our clustering tracers
is crucial.  The general result (an additional step with respect
to galaxies), is that our fiducial low-$\Omega_M$ CDM model
is capable to match very well {\bf both} the shape and amplitude of
the cluster \pk \cite{reflex_pk}.   

This shape agrees well (yet with a different, unknown bias) also with
the power spectrum of the distribution of QSO's from the 2QZ survey, a
large redshift survey of colour-selected quasars, also based on the
2dF spectrograph at the Anglo-Australian Telescope \cite{2dF_QSO}.

As can be seen from fig.~\ref{fig:pk}, the low-$\Omega_M$ CDM model
predicts a maximum for \pk around $k=0.02 \kmpc$. 
One important meaning of this turnover (which is an imprint of the
horizon size at the epoch of matter-radiation equality \cite{Paddy})
is that of an 
``homogeneity scale'', above which (smaller $k$'s) the
variance drops below the 
Poissonian value.  In a pure fractal Universe, for example, \pk would
continue to rise when moving to smaller and smaller $k$'s
\cite{Guzzo97}. In fact, 
at least visually the data of Fig.~\ref{fig:pk} do not really show a
convincing indication for a maximum.  In addition, on such extremely
large scales ($\lambda > 500 \hmpc$), the effect of the survey 
geometry on the measured power can be very significant, resulting in
an effective survey {\sl window function} in Fourier space which is
convolved with the true underlying spectrum (e.g. \cite{Tegmark_2dF}).
For highly 
asymmetric geometries, the plane-wave approximation intrinsic in the
Fourier decomposition fails, and the convolution with the window
function easily mimics a turnover in a spectrum with whatever shape (e.g
\cite{pk_esp}).  
The best solution in such cases is to resort to
survey- and clustering- specific eigenfunctions as those provided by
the Karhunen-Loeve (KL) transform \cite{Vogeley_Szalay}.  An application of
this technique to the REFLEX data \cite{KL_Peter} seems to confirm the
reality of a turnover at $k \simeq 0.023 \kmpc$, consistent with a CDM 
shape parameter $\Gamma = 0.14^{+0.13}_{-0.07}$, essentially the same
that best describes the current 2dF and CMB data
\cite{Efstathiou2001}, corresponding to an $\Omega_M\simeq 0.2$
CDM model. 

The SDSS will provide unique information around the scale of the
expected peak of $P(k)$, in particular through the Luminous Red Galaxy
(LRG) sample.  A first application of the KL transform to very early SDSS
angular data is presented in \cite{Szalay_KL_SDSS}, where it is shown
that even from only 222 square degrees, the survey is able to
sample the peak of the power spectrum.  A parallel analysis using
a different technique \cite{Dodelson_SDSS_pk} finds a best-fitting CDM
spectral shape 
$\Gamma=0.14^{+0.11}_{-0.06}$, i.e. virtually the same as measured by
2dF and REFLEX, again indicating an impressive convergence
of independent observations towards the same low-$\Omega_M$ CDM model.

The SDSS LRG sample will also be useful to verify the presence of 
{\sl baryonic features}, in $P(k)$, due to oscillations in 
the baryonic component within the last-scattering surface
\cite{Eisenstein_Hu}. A possible detection has been claimed in an
analysis of some cluster and galaxy samples previous to 2dF and REFLEX
\cite{Miller_peaks}.  These features are expected to be of very low
amplitude\footnote{Remember, however, that 
these considerations are strictly valid for the matter power spectrum,
not for the {\sl biased} field represented by galaxies or clusters for
which, in principle, one could speculate that such features might be
non-linearly amplified.}, unless  
the baryon density is 
much higher than currently established.  In fact, similar wiggles seen
in the 2dF power spectrum are interpreted as an artifact of the survey
window function \cite{Percival2001}, while the REFLEX data do not show
convincing evidence so far.

\subsection{The Two-Point Correlation Function}

\begin{figure}
\centering
\epsfxsize=8cm 
\epsfbox{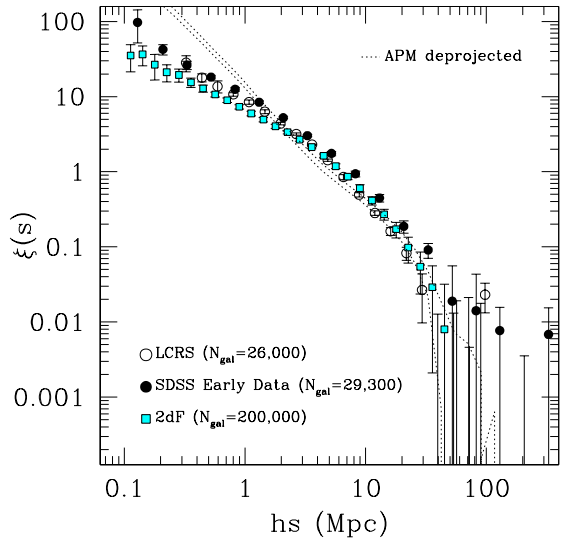} 
  \vspace{-1.5cm}
%
\caption{Estimates of the galaxy two-point correlation
function from the new 2dF \cite{Hawkins}
and SDSS \cite{Zehavi} data, compared to the previous Las Campanas
Redshift Survey \cite{LCRS_xi}. The
dotted lines show instead a correlation function in {\sl real space},
obtained through deprojection from the APM 2D galaxy catalogue \cite{Baugh96}
under two different assumptions about galaxy clustering evolution.
}
\label{xi-surveys}
\end{figure}
In fact, the simplest statistics one can compute from the data and
also that for which the selection function is more directly
corrigible, is not
the power spectrum, but rather its Fourier transform, the {\sl
two-point correlation function} $\xi(r)$, which measures the
excess probability over random to find a galaxy at a separation $r$
from a given one \cite{Peebles80}. Fig.~\ref{xi-surveys} shows
the correlation function measured in {\sl redshift space}, $\xi(s)$
(see next section for definitions), from the 2dF and SDSS current data sets
\cite{Hawkins,Zehavi}, compared to the previous LCRS \cite{LCRS_xi}.  Also shown (dotted lines) is
$\xi(r)$ reconstructed from the APM angular survey \cite{Baugh96}.

The figure shows that, especially for the two newest surveys, the shape
of $\xi(s)$ is extremely well described 
between 0.1 and $50\hmpc$, by a power-law form $\sim
(s/s_o)^{-\gamma}$, with a {\sl correlation length} $s_o\simeq
8\hmpc$.  The overall difference with the $\xi(r)$ from the
APM survey (which is in real space, being based on a deprojection of
angular clustering), is mostly due to redshift-space effects, that I
will address in detail in the next section.   Note how $\xi(s)$ maintains a
low-amplitude, positive value out to separations 
of more than $50\hmpc$, with the 2dF and SDSS data possibly implying a 
zero-crossing scale approaching $100\hmpc$.  This comparison shows
explicitly why large-size galaxy surveys are so important, given the
weakness of the clustering signal at these separations\footnote{There
is quite a bit of confusion in technical papers on the term ``scale''
when comparing results from power spectra and correlation function
analyses.  A practical ``rule of thumb'' which works about right with
well-behaved spectra is that a scale $k$ in the power spectrum,
corresponding to a spatial wavelength $\lambda=2\pi/k$, relates
approximately to  $r\sim \lambda/4$ in $\xi(r)$.}.

\subsection{Velocity Distortions in the Redshift-Space Pattern}

The separation $s$ between two galaxies computed using their observed
redshifts is not a true distance: the red-shift observed in the galaxy
spectrum is in fact the quantity $cz=cz_{\rm
cosmological}+v_{\rm pec_{||}}$, where $v_{\rm pec_{||}}$ is the
component of the 
galaxy peculiar velocity along the line of sight.   This contribution
is typically of the order of $100\, \kms$ for galaxies in the general
field, but can rise above $1000\, \kms$ within high-density regions
as rich clusters of galaxies.
Fig.~\ref{xi-surveys} shows explicitly the consequence of such {\sl
redshift-space distortion} for the correlation function: $\xi(s)$ (all
points) is {\sl flatter} than its real-space counterpart (dotted
lines).  This is the result of two  
concurrent effects: on small scales ($r\mincir 2\hmpc$), clustering
is suppressed by high 
velocities in clusters of galaxies, that spread close pairs along the
line of sight producing what in redshift maps are sometimes called
``Fingers of God''.  
Some of these are perhaps recognisable in
Fig.~\ref{2dF_cones} as thin radial structures.
On the other hand, large-scale coherent streaming flows of galaxies
towards high-density structures enhance the apparent contrast of
these, when seen perpendicularly to the line of sight. This effect, on
the contrary, amplifies $\xi(s)$ above $\sim 3 - 5 \hmpc$, as evident in
Fig.~\ref{xi-surveys}.

\begin{figure}
\centering
\epsfxsize=7.5cm 
\epsfbox{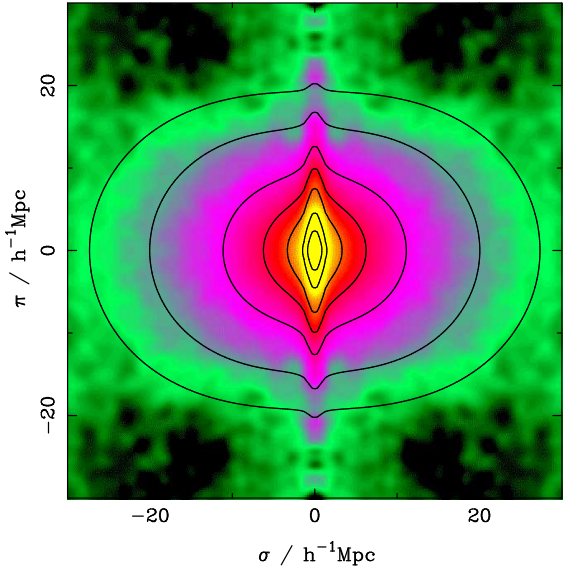} 
  \vspace{-1cm}
\caption{The bi-dimensional correlation function \xip (with $r_p$
called instead $\sigma$) from the 2dF redshift survey.  The
large-scale deviation from circular symmetry is a measure of the level
of infall of galaxies onto superclusters, proportional to
$\beta=\Omega^{0.6}/b \simeq 0.43 $ \cite{2dF_xipz}. 
}
\label{2dFbutterfly}
\end{figure}
Such peculiar velocity contribution can be disentangled by
computing the two-dimensional correlation function $\xi(r_p,\pi)$,
where the separation vector $s$ between a pair of galaxies is decomposed
into two components, $\pi$ and $r_p$, parallel and perpendicular to
the line of sight respectively (see \cite{Hamilton} for details).
%
%
The result is a bidimensional map, whose iso-correlation contours look as
in Fig.~\ref{2dFbutterfly}, where \xip computed for the 2dF survey is
plotted \cite{2dF_xipz}. 

Redshift-space distortions might seem only an annoying feature, as
they hide the true clustering pattern from direct investigation.  In
fact, they contain important information as galaxy motions are a 
direct dynamical probe of the mass distribution
\cite{Kaiser_dist}.  Non-linear distortions are
a measure of the ``temperature'' of the galaxy soup on small scales,
and they are in principle related to $\Omega_M$ through a {\sl Cosmic
Virial Theorem} \cite{Peebles80}, which however has been shown to be
difficult to apply in practice to real data \cite{Fisher94b}.  
Linear distortions produced by infall provide a way to measure the parameter
$\beta=\Omega_M^{0.6}/b$, i.e. essentially the mass density of the
Universe modulo the bias factor.  As thoroughly explained in the excellent 
review by Andrew Hamilton \cite{Hamilton}, this can be achieved by
measuring the oblate compression of the contours of $\xi(r_p,\pi)$
along $\pi$.  One way to do this is to expand $\xi(r_p,\pi)$ in
spherical harmonics.  In linear perturbation theory, only the monopole
$\xi_0(s)$, quadrupole $\xi_2(s)$ and hexadecapole $\xi_4(s)$ are
non-zero, and $\beta$ can in principle be derived directly through the
following ratio, which should be independent of $s$ \cite{Hamilton}
\begin{equation}
{\xi_2(s) \over {\xi_0(s) - \bar \xi_0}(s)} = {{{4\over 3}\beta +
{4\over7}\beta^2}\over 
{{1 + {2\over3}\beta + {1\over5}\beta^2}}}\,\,\,\,\,,
\end{equation}
where $\bar \xi_0(s) = 3s^{-3}\, \int_0^s \xi_0(x)x^2\,dx$ is the averaged correlation function within the
radius $s$. In practice, linear and non-linear effects are interlaced out to
fairly large scales ($\sim 20$ h$^{-1}$ Mpc), and require a careful
modeling.  This has been done, using a simple but effective
phenomenological approach, for the 2dF correlation function of
Fig.~\ref{2dFbutterfly}, producing one of the most remarkable results
of past year \cite{2dF_xipz}; the 2dF quadrupole-to-monopole ratio is
best reproduced\footnote{Since the conference, a more sophisticated
error analysis was applied to the 100,000 redshift public release
\cite{Tegmark_2dF}, obtaining essentially the same 
value of $\beta$, but with a 1-$\sigma$ error of $\pm 0.16$.}  
by a model with
$\beta=0.43\pm 0.07$.  If 2dF galaxies are unbiased ($b\simeq
1$), this would imply $\Omega_M\simeq 0.25$.

\begin{figure}
\centering
\epsfxsize=7.5cm 
\epsfbox{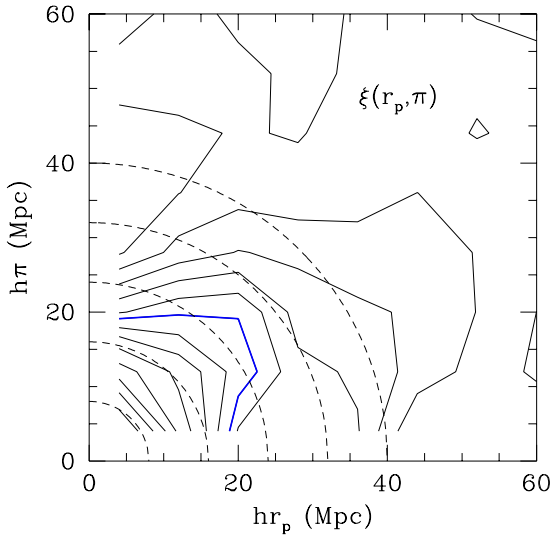} 
 \vspace{-1cm}
\caption{\xip from the REFLEX survey of X-ray clusters of galaxies
(here limited to the first quadrant only, where all the 
information is contained). Note the compression of the contours along
the redshift 
($\pi$) direction, evidence of significant streaming velocities
towards high-density regions, and the lack of any stretching at very
small $r_p$'s (there are no ``Fingers of God'' made by clusters!)
\cite{reflex_xi,Guzzo_csipz_REFLEX}.}
\label{fig:reflex_csipz}
\end{figure}

Clusters of galaxies clearly also partake in the overall motion of
masses produced by cosmological inhomogeneities.  Line-of-sight
spurious effects (as projections in optically-selected cluster
catalogues \cite{Collins95} or large redshift errors) and
limited statistics, prevented so far the detection of true velocity
anisotropies in cluster \xip maps.  Fig.~\ref{fig:reflex_csipz} plots
\xip for the REFLEX survey. Here we have a clear indication of
compression of the contours along the line of sight, of the kind
expected by the linear infall of clusters towards superstructures.
The statistical signature of cluster motions seems therefore to have
been detected, and at the time of writing accurate experiments using
large mock realisations of the REFLEX survey are under way, to
quantify precisely the confidence level at which $\beta$ can be
estimated from this map \cite{reflex_xi,Guzzo_csipz_REFLEX}.

\section{STRUCTURE AT HIGH REDSHIFTS AND GALAXY FORMATION}

The final part of my talk was dedicated to a quick overview of the
status of large-scale structure studies at $z\magcir 1$  and
its intimate connection with the formation of galaxies themselves.  It
is practically impossible with the limited space available here to
give a fair account of the fervent activity in the field of galaxy
formation and evolution. For this, the reader is addressed to
other more specific reviews available in the literature
(e.g. the excellent lectures by Ellis in \cite{Ellis2001}).  Here I
will limit the discussion to some specific points concerning the
possibility to trace the evolution of structure back in time using
galaxy redshift surveys.

\subsection{Evolution of Clustering}

Ideally, studying the evolution of structure through deep ($z>0.5$) redshift
surveys would provide a further powerful way to constrain
models and in particular to measure cosmological parameters as
$\Omega_M$, which governs the growth rate of structures.  This ideal
dream is in practice hampered by two factors.  First, in a classical
magnitude-limited survey the {\sl k-correction}\footnote{That is, the
fraction of light lost because of the red-shifting and stretching of
the spectrum at larger and larger distances} depends on the spectral
type, thus making the relative percentage of morphological types be a
function of redshift.  For example, when observing in a blue band the
k-correction for elliptical galaxies (which have a red spectrum) grows
much more rapidly than for young blue galaxies as spirals or
irregulars.  Since we know that locally red galaxies are more
clustered 
than blue galaxies (e.g. \cite{Zehavi}, just because of this effect
one would measure a fainter clustering at high z, independent of 
the growth rate of structure.  Secondly, and additionally, galaxies do
evolve and at large redshifts even a population selected, e.g., 
with the same rest-frame colour and luminosity range as locally
(i.e. properly k-corrected) would include galaxies which are very
probably not the progenitors of a local survey like 2dF and SDSS.
Also in this case, we would not be looking at the same
tracers at different redshifts, and so the observed evolution of the
correlation function or power spectrum would be impossible to connect
to the growth rate of fluctuations.  This introduces a profound
difference with ``local'' surveys, where the look-back time is small
compared to Hubble time: deep galaxy redshift surveys trace at the
same time the growth of the mass skeleton of the Universe and the
formation and evolution of the stellar population in the galaxies
themselves. 

\begin{figure}
\centering
\epsfxsize=6.5cm 
\epsfbox{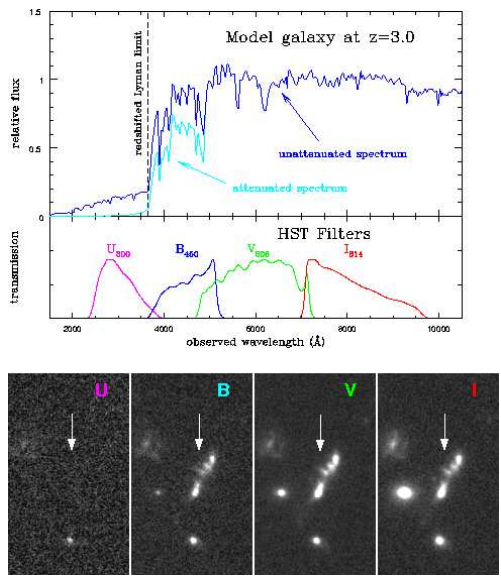} 
 \vspace{-0.5cm}
\caption{How to select ``a priori'' candidate galaxies at $z=3$: in
this figure by Marc Dickinson \cite{Dickinson_BVRI}, the
band-dropout technique is pedagogically outlined.  The upper panel shows the
spectrum of a star-forming galaxy at $z=3$, and the 
wavelength position of four filters covering the UV-visible range.
The corresponding galaxy images in the four bands are displayed in the
bottom CCD frames, with the object disappearing in the U band due to
the almost complete absorption below the redshifted Lyman limit.
}
\label{Ly-break}
\end{figure}
These difficulties are reflected by the lack of a consistent trend in the
evolution of the correlation length measured from available deep
surveys, that show values scattered between 1.8 and 5 $\hmpc$ for
redshifts in the range [0,1] (see Table 1 in \cite{Steidel98} for a
review).  
The 
large variance among different samples reflects the problems outlined
above, as the use of different spectral bands (e.g. red-selected or
blue-selected surveys), exacerbated by the limited size of surveyed
areas ($\sim 100$ arcmin), due to the time needed to collect a
spectrum for $z\simeq 1$ galaxies.  It is probably for these reasons
that most progress in the area of the evolution of clustering has come
during the last few years from 2D surveys, measuring angular
clustering over fairly large angles thanks to wide-angle CCD cameras
(see e.g. \cite{CFDF} and references therein). 

 This situation is therefore yet another example in the long list of
``cosmological probes'' which rather than cosmology turned out to be
testing evolution of a particular class of objects (were they either
radio galaxies, first-ranked cluster galaxies, or generic galaxies).
This shifts the attention in deep surveys from just using galaxies as
``test particles'' 
to understanding the way their baryonic component was assembled and
evolved as to shape today's colours and morphologies
\cite{Ellis2001,Steidel98}.  There are therefore good reasons for
going deeper with redshift surveys: the picture becomes probably more
complicated to interpret cosmologically, but certainly richer in
astrophysical details.  ``Big'' questions that need investigation at
high $z$ are, for example, (1) do we see evidence of continuous
merging, as expected in the ``standard'' hierarchical model?  (2) Can
we reconcile with this scenario the observation of a consistent
fraction of old massive elliptical galaxies at $z>1$? (3) How
the assembly of galaxy masses translates into the star formation
history of the Universe?  (4) How do we connect this to the ionization
history of the Universe? (5) How did galaxy morphologies (and their
relation with local density) originate? \cite{Ellis2001}

\subsection{Clustering at $z> 1$: High-bias Objects}

The best current flux-limited deep surveys reach $z\sim 1$
\cite{CFRS,CNOC}. However, if the goal is to select samples at very
large redshift, moving the 
flux (magnitude) limit to even fainter values is not the most
efficient option (although it guarantees a higher level of control
over selection effects).   Alternatively, an important advance of the
last few years has been the ability to select through photometric
means (i.e. using CCD images in different filters), almost
volume-limited samples of galaxies with mean redshifts $z\simeq 3$ and
$\simeq 4$ \cite{Steidel98,Steidel_Hamilton}.  This is an efficient
alternative to pure flux-limited samples, which even at very faint
magnitude limits are dominated by a bulk of faint $z<0.5$ objects.
The technique is based on the detection of specific features
(``breaks'') in galaxy spectra using multi-band imaging through
appropriate sets of filters.  The most notable of such features is the
Lyman break at 912 \AA\, which is prominent in star-forming galaxies.
At $z=3$, the break is redshifted to 3650 \AA, thus falling between the
so-called $U$ and $B$ filters.  For this reason, a star-forming $z\sim
3$ galaxy 
will be detected in $B$, but will be almost invisible in $U$ (see
Fig.~\ref{Ly-break}).  A similar reasoning can be applied to hunt for
$z\sim 4$ galaxies, using instead the $B$ and $V$ filters.  A fairly
large sample ($\sim 1000$ galaxies) at $z=3$ has been constructed
in this way during the 1990's by Steidel and collaborators,
spectroscopically confirmed through intense use of the Keck 10-m
telescopes \cite{Steidel_Hamilton}.

Ly-break galaxies turn out to be {\sl highly biased} objects.  In fact,
they show a very strong clustering, with a correlation length
comparable to that of normal galaxies at the present epoch
\cite{Giavalisco}. This implies that they cannot be representative of
the mass clustering at that epoch.  A consistent explanation is that
they are the progenitors of massive elliptical galaxies that today
populate rich clusters of galaxies, undergoing their most active phase
of star formation \cite{Governato_Nature}. 

\subsection{Future Deep Redshift Surveys}

Although custom-selected samples of high-redshift objects as
Lyman-break galaxies are certainly fundamental for understanding
specific scientific issues, it is difficult to use them to trace the
evolution of the overall population.  To make a further major step in
our understanding of how structure and galaxies formed and evolved, we
need new surveys which: (1) Are based on multi-band photometric
information, such that selection effects can be understood as finely
as possible and specific morphologies and rest-frame colours can be
traced back in time over a sensible range; (2) Cover areas which are
wide enough to reduce ``cosmic variance''; (3) Provide, a combination
of depth, volume and statistics (number of objects) comparable to
local ongoing surveys as 2dF and SDSS.

There are currently two ambitious projects which are due to start
during 2002 and that are expected to be close to these {\sl
desiderata}. One is the DEEP2 survey \cite{DEEP}, that will collect
spectra for $\sim$60,000 galaxies between $z\sim 0.7-1.5$, over four
$2^\circ$ by $0.5^\circ$ strips of sky using the Keck 10~m telescope.
The other is the VIRMOS survey, that will similarly observe
$\sim$150,000 galaxy spectra using mainly the new VIMOS spectrograph at
the VLT 8~m telescope, splitted into a ``wide'' survey over $\sim 16$
deg$^2$ to a red magnitude $I_{AB}=22.5$ ($\sim 100,000$ redshifts),
plus a ``deep'' survey over $\sim 1$ deg$^2$ to $I_{AB}=24$ ($\sim
50,000$ redshifts), in addition to one or more ``ultra-deep'' probes
to $I_{AB}=25$ over some  arcminute-sided fields, using an Integral
Field Unit \cite{VIRMOS}.

\section{SUMMARY}

In conclusion, I hope to have passed the sensation that we do live in a
glorius time for cosmology, which has 
finally become a mature science, not anymore an entertaining playground
for almost pure speculation.  In fact, we never had such a wealth of
data at our disposal, by which we are pinning down the values of
cosmological parameters to high accuracy
(e.g. \cite{CMB}), while at the same time being able to
study galaxy formation almost in the act, thanks to new powerful
telescopes.  The few observational facts we have reviewed here
contribute to further reinforce the remarkable convergence of
different observables (CMB, large-scale structure, distant
Supernovae, cluster evolution, to mention a few) towards what we
called in the introduction the current standard model, i.e. one with a
flat geometry ($\Omega_{total}=1$), apparently guaranteed by the
combination of a dominating Cold Dark Matter component
($\Omega_{CDM}\simeq 0.3$, $\Omega_{baryon}\simeq 0.02$) and a Dark
Energy of unknown nature (the cosmological constant,
$\Omega_{\Lambda}\simeq 0.7$ ).  Several
interesting talks at the TAUP2001 meeting were devoted to the
detection of baryonic and non-baryonic dark matter.  I invite
the reader to look at the corresponding contributions in this volume,
however, and 
test whether he/she is not left with some uneasiness as our wonderful
``standard'' cosmological model seems in fact to be so far essentially
based on a) a {\it Dark Matter} we do not detect; b) a {\it Dark
Energy} we do not understand; c) a fraction of Baryons we cannot
completely find!  Yet everything seems to work: isn't this reminiscent
of epicycles? 

{\bf Acknowledgments.} I thank the organizers of TAUP2001 for inviting
me to give this review, and in particular Alessandra Di Chiarico for
her patience in waiting for this contribution. I am grateful to all my
collaborators without whom many of the results discussed here would
not be reality, in particular P. Schuecker, C. Collins and
H. B\"ohringer, and to A. Fernandez-Soto for a careful reading of the
manuscript and useful discusssions.  I thank I. Zehavi and E. Hawkins
for providing their clustering results in electronic form.

\end{document}